\documentclass[a4paper,11pt]{article}
\usepackage{pos}

\usepackage{xcolor}      % for coloured text
\usepackage{graphicx}    % for importing pictures
\usepackage{csquotes}    % for \enquote
\usepackage{tikz}        % for words on top of slides
\usepackage{braket}      % for bra-ket notation
\usepackage{placeins}    % For the \FloatBarrier command - used manually
\usepackage{soul}        % Something to do with hyphenation

\newcommand{\tilm}[1]{\skew{1.5}\widetilde{#1}}

\DeclareRobustCommand{\order}[1]{\mathcal{O}\left(#1\right)}

\newcommand{\rb}[1]{\!\left(#1\right)}
\newcommand{\sq}[1]{\left[#1\right]}

\DeclareRobustCommand{\eqnr}[1]{Eq.~$\left(\ref{#1}\right)$}

\newcommand{\Fig}[1]{Fig.~\ref{#1}}
\newcommand{\Tab}[1]{Table \ref{#1}}

\newcommand{\Refl}[1]{Ref.~\cite{#1}}    %Might need to make more often than usual
\newcommand{\Refls}[1]{Refs.~\cite{#1}}  
\title{$U(1)_A$ symmetry restoration at finite temperature with mesonic correlators}
\author*[a]{Ryan Bignell}
\author[b]{Gert Aarts}
\author[b]{Chris Allton}
\author[c]{Benjamin J\"ager}
\author[d]{Seyong Kim}
\author[e,a]{Jon-Ivar Skullerud}
\author[b,f]{Antonio Smecca}

\affiliation[a]{School of Mathematics \& Hamilton Mathematics Institute,
  Trinity College, Dublin, Ireland}

\affiliation[b]{Centre for Quantum Fields and Gravity, Department of Physics,
  Swansea University, Swansea, SA2 8PP, United Kingdom}

\affiliation[c]{Quantum Theory Center ($\hbar$QTC) \& Danish IAS, Department of Mathematics and Computer Science,\\
  University of Southern Denmark, 5230, Odense M, Denmark}

\affiliation[d]{Department of Physics,
  Sejong University, Seoul 143-747, Korea}

\affiliation[e]{Department of Physics,
  Maynooth University--National University of Ireland Maynooth, County Kildare, Ireland}

\affiliation[f]{INFN,
  Sezione di Roma Tre, Via della Vasca Navale 84, I-00146 Rome, Italy}

\emailAdd{bignellr@tcd.ie}
\emailAdd{\{g.aarts,c.allton\}@swansea.ac.uk}
\emailAdd{jaeger@imada.sdu.dk}
\emailAdd{skim@sejong.ac.kr}
\emailAdd{jonivar.skullerud@mu.ie}
\emailAdd{antonio.smecca@roma3.infn.it}

\abstract{The $U(1)_A$ symmetry of the massless QCD Lagrangian is explicitly broken in the quantised theory by the anomaly. It may be effectively restored at some finite temperature, which would have important consequences for the order of the chiral transition and the QCD phase diagram. It has been argued in the literature that one way to probe the effective restoration of $U(1)_A$ is to check for the degeneracy of pseudoscalar and flavour non-singlet scalar correlators. In this work, we consider a new method of examining this degeneracy based upon hadron correlation functions on the anisotropic FASTSUM ensembles. The anisotropic nature and our newest Generation 3 ensembles aid in a determination of the effective restoration of the $U(1)_A$ symmetry which we find to be $T_{U(1)_A} \sim 320$ MeV, well above the chiral transition temperature, which is $T_{\rm pc} \sim 180$ MeV for our choice of Wilson-Clover fermions.}%which we find to be $T_{\text{chiral}} <T_{U(1)_A}< 2\,T_{\text{chiral}}$.}

\FullConference{The 42nd International Symposium on Lattice Field Theory (LATTICE2025)\\
2-8 November 2025\\
Tata Institute of Fundamental Research, Mumbai, India\\}

\begin{document}
\maketitle
\section{Introduction}
%\vspace{-0.015\textheight}
While the massless QCD Lagrangian is invariant under the symmetry $SU(n_f)_L \times SU(n_f)_R \times U(1)_V \times U(1)_A$ (for $n_f$ flavours), the $SU(n_f)_L \times SU(n_f)_R$ \textit{chiral} symmetry is spontaneously broken at zero temperature and the $U(1)_A$ symmetry is explicitly broken by quantisation~\cite{Bell1969,Adler1}. The chiral symmetry is restored at some finite temperature, while the $U(1)_A$ symmetry is thought to be effectively restored --- it remains explicitly broken but the breaking effects are suppressed~\cite{Nicola:2020smo}. The pseudo-critical temperature for chiral symmetry restoration is well known to be about $T_{\rm pc} \simeq 154$ MeV~\cite{HotQCD:2018pds,Borsanyi:2020fev,Gavai:2024mcj} for physical quark masses. In contrast, the effective restoration of the $U(1)_A$ symmetry is still not settled~\cite{Lahiri:2021lrk,Borsanyi:2025ttb}. It is becoming evident that $U(1)_A$ symmetry is likely not (effectively) restored at the chiral transition temperature, but where and if the restoration occurs is yet unclear~\cite{Brandt:2016daq,Aoki:2020noz,Kaczmarek:2021ser,Dentinger:2021khg}.
\par
The restoration of $U(1)_A$ symmetry should be manifest~\cite{Shuryak:1993ee,Cohen:1996ng,Cohen:1997hz} at the level of mesonic correlation functions, specifically through the degeneracy of the pseudoscalar $(\pi)$ and the flavour non-singlet scalar meson $(\delta)$ channels. This can be examined through the spectral density of the Dirac operator~\cite{Dick:2015twa,Aoki:2012yj}. Due to our choice of Wilson-Clover fermions~\cite{Sheikholeslami:1985ij}, which makes direct spectral density studies difficult, we instead consider the correlation functions themselves.
%\tcr{\\SEE 3.1.3 (page 17) of 2512.08843 for inspiration~\cite{Borsanyi:2025ttb}}
\par
The most common approach using mesonic correlators is to consider the susceptibilities
\begin{align}
  \chi_\pi = \sum_{\vec{x},\tau}\,G_\pi(\vec{x},\tau) \quad\quad \text{and} \quad\quad \chi_\delta = \sum_{\vec{x},\tau}\, G_\delta(\vec{x}, \tau),
  \label{eqn:def1}
\end{align}
where $G_{\sq{\pi,\delta}}(x)$ are the pseudoscalar and flavour non-singlet scalar two-point correlation function respectively. The difference in susceptibilities $\chi_\pi - \chi_\delta$ is used to study the restoration. These susceptibilities are used in an attempt to remove ultraviolet divergences~\cite{HotQCD:2012vvd,Suzuki:2019vzy}, which even good chiral actions such as Domain Wall Fermions suffer from~\cite{Cossu:2013uua}. In order to further reduce the effect of these ultraviolet -- or short distance -- artefacts we introduce two additional methods. In combination with the high-temperature anisotropic \textsc{Fastsum} ensembles, these enable an examination of where $U(1)_A$ symmetry is effectively restored at  high temperature.
%\vspace{-0.015\textheight}
\section{Ensembles}
%\vspace{-0.015\textheight}
\begin{table}[b]
\centering
%\begin{tabular}{r|llllllll}
%$N_\tau$         & $256$ & $128$ & $112$ & $96$  & $80$  & $76$  & $72$ &  \\
%  $T (\text{MeV})$ & $50$  & $100$ & $114$ & $133$ & $160$ & $169$ & $178$ & \\ \hline
%  $N_\tau$         & $68$  & $64$  & $56$  & $48$  & $40$  & $36$  & $32$  & $24$  \\
%$T (\text{MeV})$ &  $188$ & $200$ & $229$ & $267$ & $320$ & $356$ & $400$ & $534$
%\end{tabular}
\resizebox{\textwidth}{!}{%
\begin{tabular}{r|
%lllllllllllllll}
rrrrrrrrrrrrrrrr}
$N_\tau$         & $256$ & $128$ & $112$ & $96$  & $80$  & $76$  & $72$  & $68$  & $64$  & $56$  & $48$  & $40$  & $36$  & $32$  & $24$  \\
$T (\text{MeV})$ & $50$  & $100$ & $114$ & $133$ & $160$ & $169$ & $178$ & $188$ & $200$ & $229$ & $267$ & $320$ & $356$ & $400$ & $534$
\end{tabular}%
}
\caption{Generation 3 ensembles showing the temporal extent $N_\tau$ and the corresponding temperature.}
\label{tab:ens}
\end{table}
In this work we use the latest \textsc{Fastsum} \enquote{Generation 3} ensembles~\cite{Skullerud:2025xva,Gen3}. These are anisotropic ensembles with the temporal lattice spacing $a_\tau$ a factor of $\xi = a_s / a_\tau \sim 7$ times smaller than the spatial lattice spacing. This allows a fine spread of temperatures using the fixed-scale approach, wherein the temperature is changed by changing the temporal extent of the lattice, $T=1/(a_\tau N_\tau)$. Following the HadSpec Collaboration~\cite{HadronSpectrum:2008xlg,Edwards:2008ja} we use an $\order{a^2}$ improved gauge action and an $\order{a}$ improved Wilson-Clover fermion action.
\par
The light quark mass is tuned to closely reproduce the pion mass of our previous \enquote{Generation 2} ensembles~\cite{Aarts:2014nba,Aarts:2020vyb,aarts_2023_8403827} giving $m_\pi =378(1)$ MeV with a near physical strange quark mass. The spatial lattice spacing is $a_s = 0.10796(57)$ fm with anisotropy $\xi = a_s / a_\tau = 7.0161(94)$. The pseudocritical temperature, measured via the chiral condensate, is $T_{\rm pc} \sim 180$ MeV. The Generation 3 ensembles span a broad range of temperatures with $T \in\sq{50,\,534}$ MeV as shown in \Tab{tab:ens}. The zero-temperature ensemble, with $N_\tau=256$, has a spatial extent of $N_s=24$, while for the others it is $N_s=32$. A full description of these ensembles will be detailed in a future publication~\cite{Gen3}. 
%\vspace{-0.015\textheight}
\section{Results using local correlation functions}
%\vspace{-0.015\textheight}
Here we discuss the three methods used to examine the degeneracy of the pseudoscalar and flavour non-singlet scalar mesons using \enquote{local-local} (point source and sink) correlators.
%\vspace{-0.015\textheight}
\subsection{Standard Definition}
%\vspace{-0.015\textheight}
\begin{figure}[ht]
  \centering
  \includegraphics[width=0.486\columnwidth]{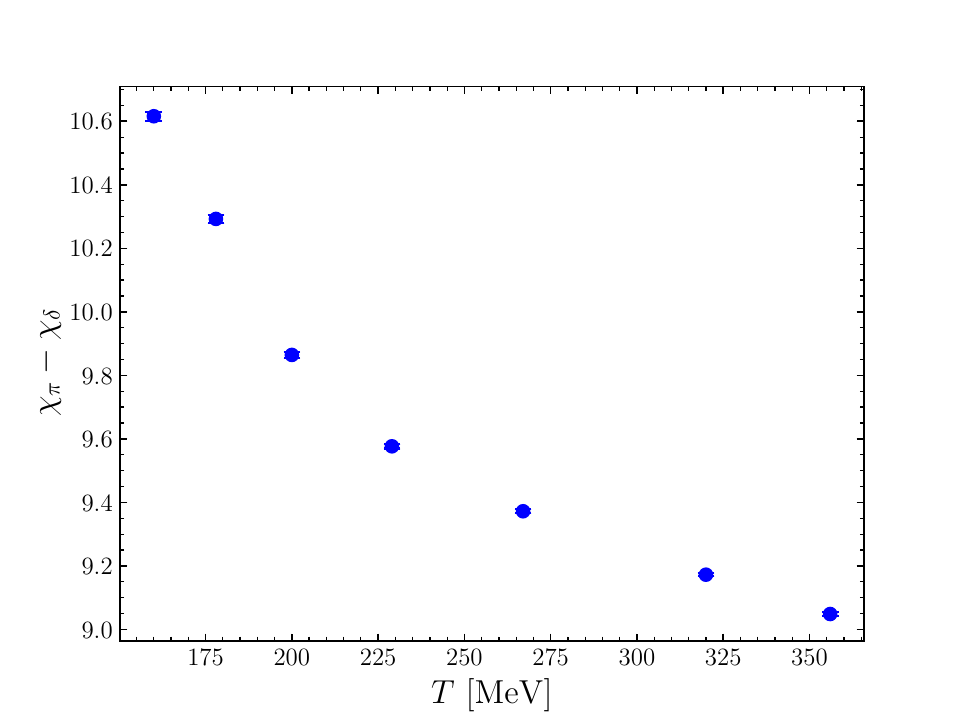} \hfill
  \includegraphics[width=0.486\columnwidth]{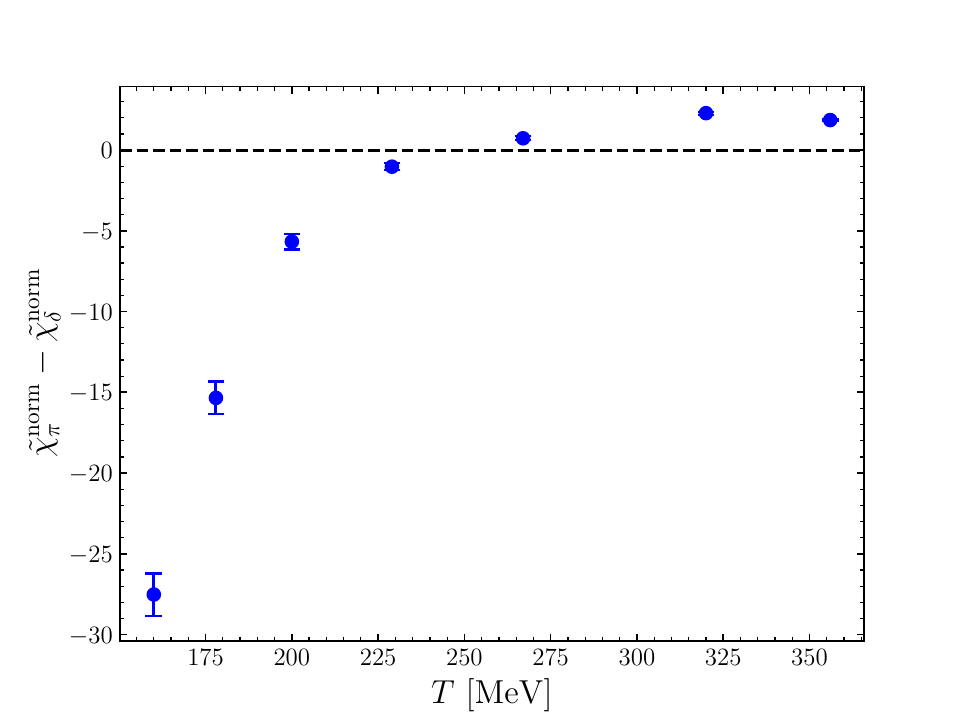}
  \caption{\textbf{Left:} Subtracted susceptibilities $\chi_\pi-\chi_\delta$, see \eqnr{eqn:def1}. \textbf{Right:} Subtracted normalised susceptibilities $\tilde\chi_\pi^{\rm norm}-\tilde\chi_\delta^{\rm norm}$, see \eqnr{eqn:def2}. Here the correlator has been normalised at the mid-point first and the sum starts at $\tau_{\text{min}}\,T = 0.2$.}
  \label{fig:def12}
\end{figure}
The results using the standard definition (\ref{eqn:def1}) are shown in \Fig{fig:def12} (left).  Although there is a clear decrease in the difference of susceptibilities as the temperature increases, the difference remains far from zero, suggesting that there is no degeneracy. This is because the operators used do not (necessarily) have the same overlap with the pseudoscalar/scalar states and hence subtraction will not show degeneracy. Furthermore, any short-distance effects due to excited states and the Wilson term are unmitigated here.
%\vspace{-0.015\textheight}
\subsection{Normalise and cutoff}
%\vspace{-0.01\textheight}
The second definition introduces a short-distance cutoff: each (temporal) correlator is normalised at its mid-point and only summed after a few time steps to reduce short-distance artefacts. We first define, for the remainder of this manuscript, $G_{\sq{\pi,\delta}}\rb{\tau} = \sum_{\vec{x}}\,G_{\sq{\pi,\delta}}\rb{\vec{x},\tau}$
%\begin{align*}
%G_{\sq{\pi,\delta}}\rb{\tau} = %\sum_{\vec{x}}\,G_{\sq{\pi,\delta}}\rb{\vec{x},\tau}
%\end{align*}
and then construct the susceptibility
\begin{align}
%\tilm{\chi}_{\sq{\pi,\delta}}^{\,\text{norm}} = \sum_{\tau_{\text{min}}}^{N_\tau}\,\frac{\sum_{\vec{x}}\,G_{\sq{\pi,\delta}}\rb{\vec{x},\tau}}{\sum_{\vec{x}}\,G_{\sq{\pi,\delta}}\rb{\vec{x},\frac{N_\tau}{2}}}.
\tilm{\chi}_{\sq{\pi,\delta}}^{\,\text{norm}} = \sum_{\tau=\tau_{\text{min}}}^{N_\tau/2}\,\frac{G_{\sq{\pi,\delta}}\rb{\tau}}{G_{\sq{\pi,\delta}}\rb{N_\tau/2}}.
  \label{eqn:def2}
\end{align}
We then take the difference $\tilm{\chi}_{\pi}^{\,\text{norm}} - \tilm{\chi}_{\delta}^{\,\text{norm}}$ as before. The improvement here is twofold -- first the dependence on the overlap of the energy states in the operators is removed by normalising with respect to the mid-point of the correlator,  where the ground state dominates, and second, short-distance effects are removed by introducing a short-distance temporal cut-off $\tau_{\text{min}}$. Hence we have enhanced the sensitivity of $\tilm{\chi}^{\text{norm}}$ to infrared effects relevant to the (effective) restoration of $U(1)_A$ symmetry. The difference in the resulting susceptibilities is shown in \Fig{fig:def12} (right). Here the sum starts from $\tau_{\text{min}}\,T = 0.2$. This difference now approaches (and crosses) zero showing (effective) $U(1)_A$ symmetry restoration. While care must be taken when choosing $\tau_{\text{min}}$, we find that in this setup $U(1)_A$ symmetry is effectively restored at $T \gtrsim 225 $ MeV.
%\vspace{-0.015\textheight}
\subsection{Normalised Ratio}
%\vspace{-0.015\textheight}
\begin{figure}[ht]
  \centering
  \includegraphics[width=0.486\columnwidth]{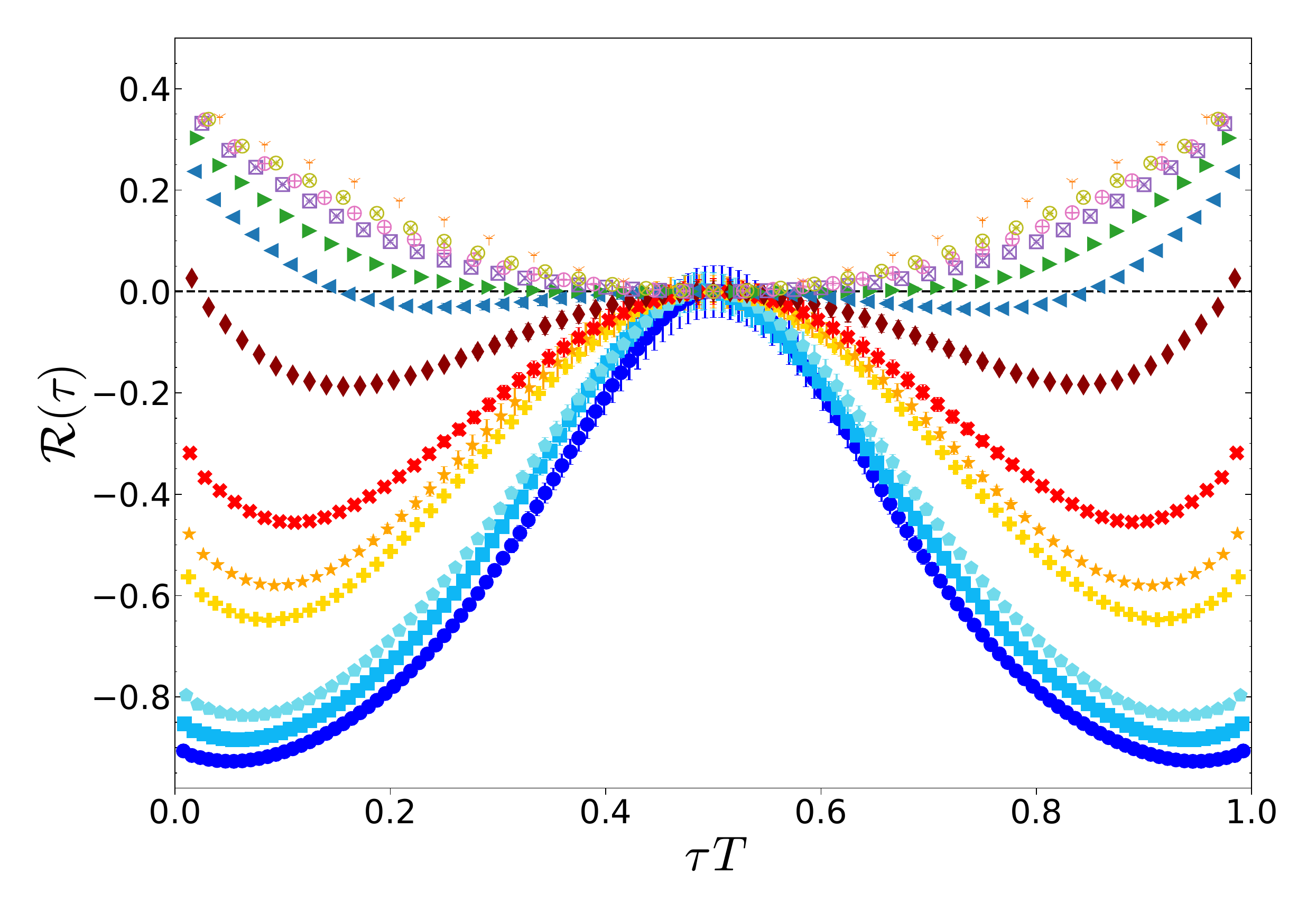} \hfill
  \includegraphics[width=0.486\columnwidth]{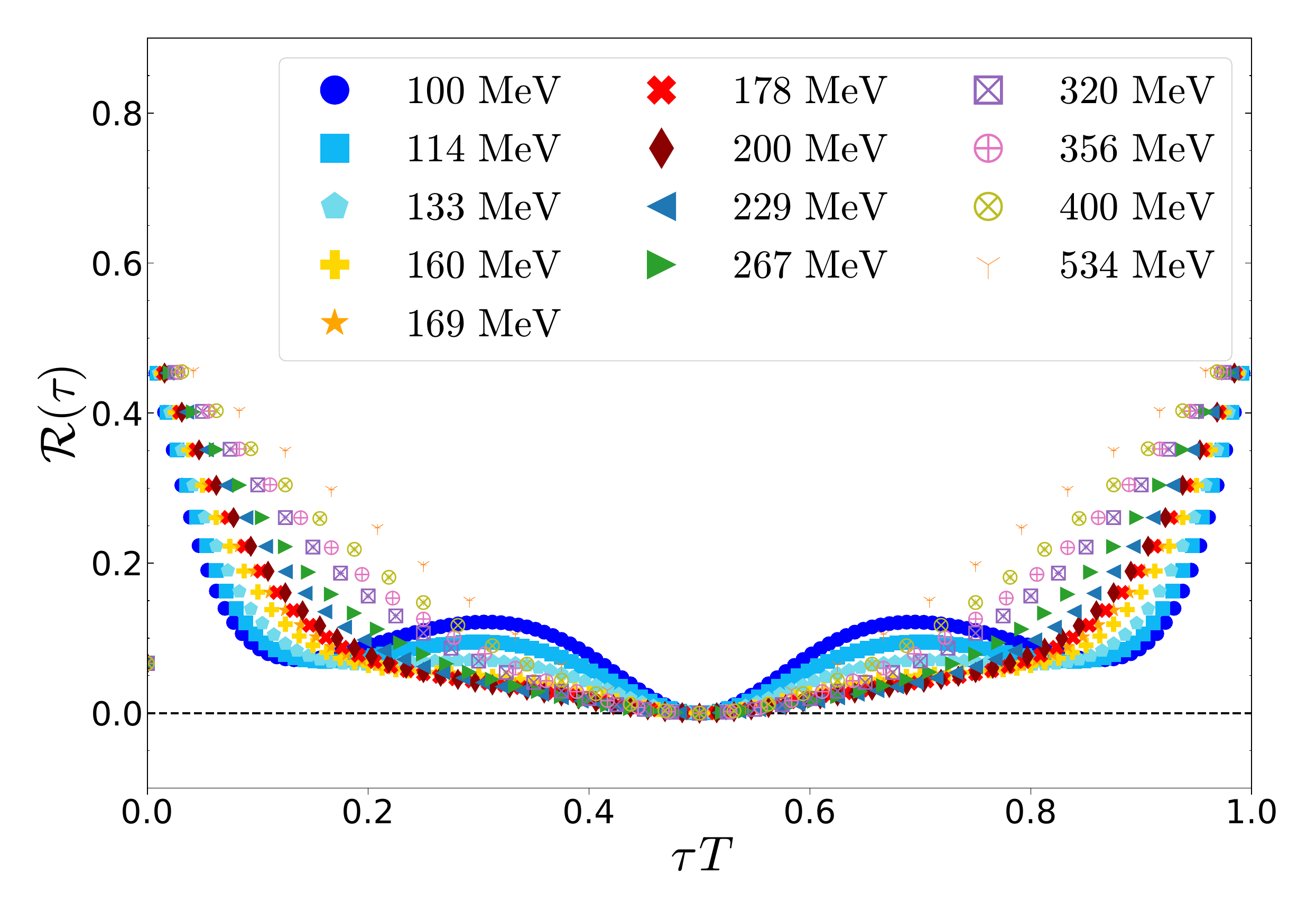}
  \caption{\textbf{Left:} Ratio of normalised correlators $\mathcal{R}\rb{\tau}$, see \eqnr{eqn:def3}. Note that the data corresponds to the legend in the figure on the right, with the temperature increasing vertically along the left-hand side of the plot. \textbf{Right:} Ratio $\mathcal{R}\rb{\tau}$ for free Wilson correlators on lattices of the same dimensions as Generation 3.}
  \label{fig:def3}
\end{figure}
The final method we consider is to construct a ratio of the difference of normalised correlators~\cite{Datta:2012fz,Smecca:2024gpu,Aarts:2020vyb}
\begin{align}
  \mathcal{R}\rb{\tau} = \frac{\tilm{G}_\pi\rb{\tau} - \tilm{G}_\delta\rb{\tau}}{\tilm{G}_\pi\rb{\tau} + \tilm{G}_\delta\rb{\tau}},
  \label{eqn:def3}
\end{align}
where $\tilm{G}\rb{\tau}$ is the mid-point normalised correlator $\tilm{G}\rb{\tau} = G\rb{\tau} / G\rb{N_\tau / 2}$. This ratio is constructed such that it is close to $\pm 1$ when the correlators are non-degenerate and the ground state masses differ substantially, and zero when they are exactly degenerate. The ratio vanishes at $\tau = N_\tau/2$ by construction. Since $m_\pi \ll m_\delta$, non-degeneracy is indicated by values close to $-1$. The mid-point normalisation handles any relative difference in the operator overlap. With different mesonic channels, this ratio has previously been used to examine the chiral transition temperature~\cite{Smecca:2024gpu,Aarts:2020vyb}
\par
This ratio is presented in \Fig{fig:def3} (left). The ratio is, of course, symmetric for mesonic correlators. It is clear that as the temperature increases, in the intermediate region of $\tau\,T \sim \sq{0.2,\,0.4}$ it gets closer to zero. At the edge of the lattice there is an upwards curvature. This behaviour can be understood through consideration of free Wilson correlators (for details, see \Refl{Smecca:2024gpu}) as in \Fig{fig:def3} (right). Here the influence of artefacts due to the Wilson term is manifest at the edges of the lattice. 
%\vspace{-0.015\textheight}
\section{Smeared Ratio}
%\vspace{-0.015\textheight}
\begin{figure}[ht]
  \centering
  \includegraphics[width=0.486\columnwidth]{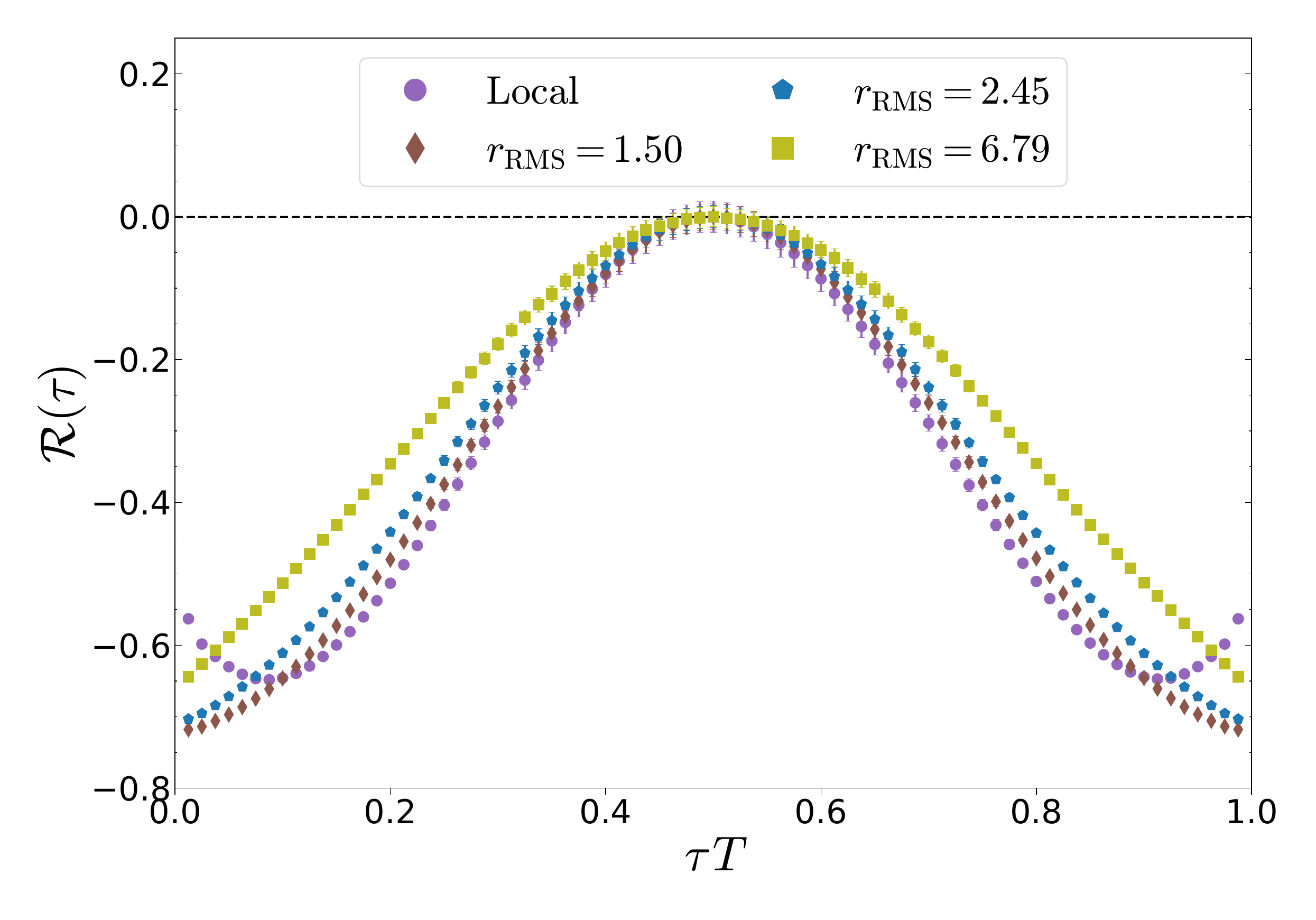} \hfill
  \includegraphics[width=0.486\columnwidth]{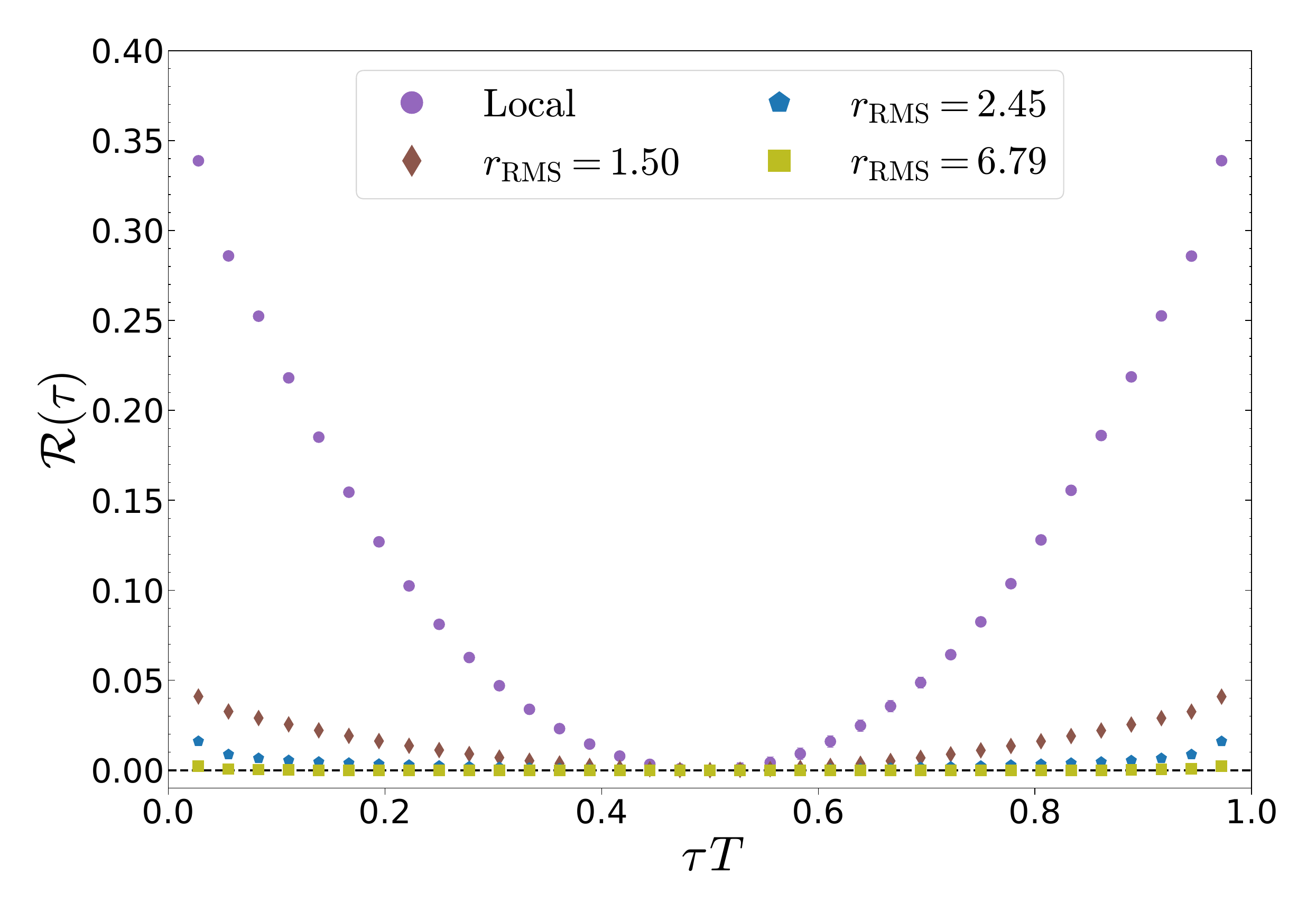}
  \caption{Ratio $\mathcal{R}\rb{\tau}$, see \eqnr{eqn:def3}, for four different smearing radii (including none) at $T\sim160$ MeV (left) and $T\sim356$ MeV (right). Note that the vertical scales are different.}
  \label{fig:smearComp}
\end{figure}
Up to now, we considered local-local correlators only. 
To reduce the impact of the artefacts due to the Wilson term, we consider now standard Gaussian smeared correlators~\cite{Gusken:1989qx,Aarts:2023nax}. Source and sink smearing is a common technique used to reduce overlap with heavy excited states, some of which are due to the nature of the Wilson-Clover action.
\par
To tune the amount of smearing used we compare the resulting ratio (\ref{eqn:def3}) for four different levels of smearing (including none) in \Fig{fig:smearComp} at two different temperatures. The amount of smearing is described by the root-mean-square radius of the Gaussian profile on a free point source. The $r_{\text{RMS}} = 6.79$ profile was chosen to match the one previously optimised for nucleon and charm baryon spectroscopy in \Refls{Aarts:2020vyb,Aarts:2023nax}; the other values were chosen to be smaller in radii.
\par
From \Fig{fig:smearComp} it is evident that at very high temperatures $\rb{T\sim356\text{ MeV}}$ that there is little difference between the smeared correlators, but that at lower temperatures $\rb{T\sim160\text{ MeV}}$ larger differences remain (note the vertical scale in the left and right plots is different). Close to the mid-point of the correlator, the ground state should be dominant and the smeared correlators should resemble the \enquote{local} or unsmeared correlator and this is observed for all but the $r_{\text{RMS}}=6.79$ correlators. Hence we discard these. Of the remaining correlators, we select the $r_{\text{RMS}}=2.45$ correlator corresponding to $\kappa=1.96\text{, }n=6$ in our smearing algorithm~\cite{Gusken:1989qx,Aarts:2023nax}, as it most effectively eliminates the upwards curve at short temporal separation.
%\vspace{-0.015\textheight}
\subsection{Integrated Ratio}
%\vspace{-0.015\textheight}
\begin{figure}[h!]
  \centering
  \includegraphics[width=0.486\columnwidth]{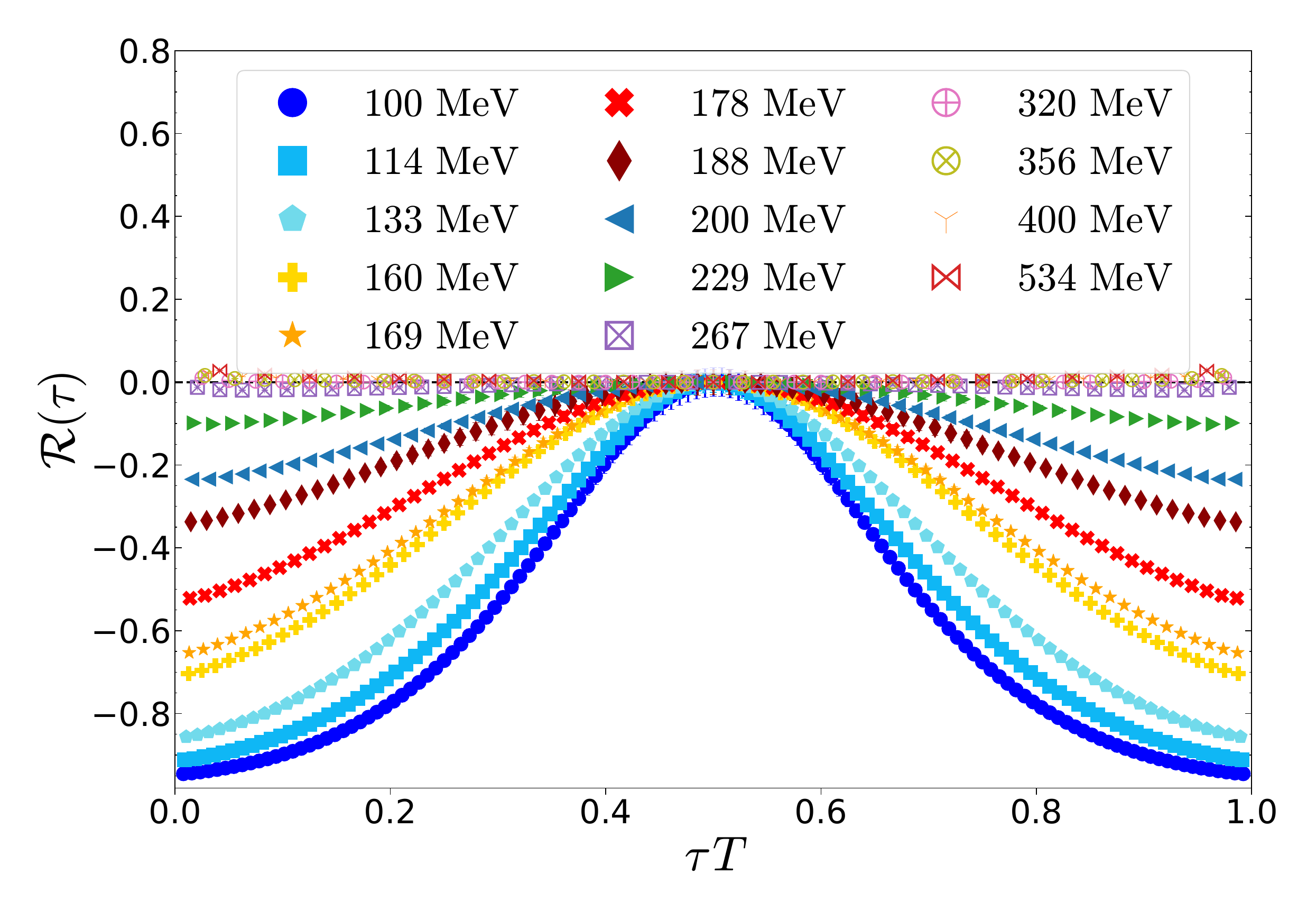} \hfill
  \includegraphics[width=0.486\columnwidth,page=26]{Figures/RexclHalf_SG_smeared.pdf}
  \caption{\textbf{Left:} Ratio $\mathcal{R}\rb{\tau}$, see \eqnr{eqn:def3}, using smeared correlators. \textbf{Right:} Corresponding integrated ratio $R$, see \eqnr{eqn:def3:int}. The orange curve is a spline interpolation used to find the intercept of the curve with zero.}
  \label{fig:def3:smeared}
\end{figure}
The ratio constructed in \eqnr{eqn:def3} is useful as it allows an examination of the influence of Wilson artefacts and the desired degeneracy (or lack thereof), but it is difficult to use it to determine at what temperature degeneracy occurs. Hence in the spirit of the original  definition (\ref{eqn:def1}) of the susceptibilities, we construct the integrated (summed) ratio
\begin{align}
  R\rb{\tau_{\text{min}},\,N_\tau /2; \,T} = \frac{\sum_{\tau_{\text{min}}}^{N_\tau / 2}\, \mathcal{R}\rb{\tau, T} / \sigma^2_\mathcal{R}\rb{\tau, T}}{\sum_{\tau_{\text{min}}}^{N_\tau / 2} 1 / \sigma^2_\mathcal{R}\rb{\tau, T}},
  \label{eqn:def3:int}
\end{align}
where $\mathcal{R}\rb{\tau, T}$ is the ratio of \eqnr{eqn:def3} and $\sigma_{\mathcal{R}}\rb{\tau, T}$ the corresponding uncertainty. As in the definition of \eqnr{eqn:def2}, there is some freedom in the choice of $\tau_{\text{min}}$ but we find that the use of smeared correlators greatly reduces this effect. This integrated ratio inherits the properties of the underlying ratio, namely that degeneracy is evident by a zero value and non-degeneracy by a value close to $-1$.
\par
In \Fig{fig:def3:smeared} (left) we show the ratio of \eqnr{eqn:def3} using smeared correlators. Note how the introduction of smearing has greatly reduced the upwards inflection at the edges of the lattice in comparison to the unsmeared correlators and that the ratio now remains close to zero once degeneracy has been reached. It is important to note that all the ratio plots are plotted as a function of $\tau\,T$ and hence an ensemble at a higher temperature has fewer points.
\par
The integrated ratio of \eqnr{eqn:def3:int}, using the smeared correlator data, is shown in \Fig{fig:def3:smeared} (right). Here the sum is taken from $\tau_{\text{min}}\,T=0.05$; hence $\tau_{\text{min}}$ is different for each temperature. The correlators are clearly not degenerate at the chiral transition temperature $T_{\rm pc} \sim 180$ MeV, but instead become degenerate at a much higher temperature. To determine the temperature at which this occurs, a cubic spline interpolation is performed and the intersection with $0$ is found. For this data, this temperature is $T_{U(1)_{A}} = 317(4)$ MeV, significantly above $T_{\rm pc}$. The uncertainty of $4$ MeV is purely statistical; a fuller determination of systematic uncertainties is planned.
\par
Although the integrated ratio becomes slightly non-zero and crosses zero after $T_{U(1)_{A}}$, this is not too concerning. Even though the smearing has mitigated artefacts due to the Wilson term, the fermion action still explicitly breaks chiral symmetry -- the impact of this seems slight in the vector/axial-vector integrated ratio as we recently examined~\cite{Smecca:2024gpu}. Our set-up is at a moderate quark mass ($m_\pi \sim 380$ MeV) and here we clearly see the effective restoration of $U(1)_A$ symmetry. 
%\vspace{-0.015\textheight}
\section{Conclusions and Future Work}
%\vspace{-0.015\textheight}
We studied the effective restoration of $U(1)_A$ symmetry on Generation 3 \textsc{Fastsum} ensembles with Wilson-Clover fermions, with a pion mass of $m_\pi \sim 380$ MeV and an anisotropy of $\xi\sim 7$, by analysing an integrated ratio of smeared $\pi$ and $\delta$ correlators.
We found that this symmetry is effectively restored at $T=317(4)$ MeV, well above the chiral transition temperature $T_{\rm pc} \sim 180$ MeV.
\par
An advantage of the anisotropic \textsc{Fastsum} ensembles is that they span a broad range of temperatures, both below and above $T_{\rm pc}$. This, combined with the anisotropic nature which enables temporal correlation functions with many points, allows a novel investigation of the susceptibilities before they are summed. In particular, the use of the (integrated) ratio has allowed for a detailed understanding of systematic effects associated with the Wilson term and the choice of $\tau_{\text{min}}$. This work confirms the lack of $U(1)_A$ restoration at $T_{\rm pc}$ and provides a new way to determine the transition temperature.
\par
In the future we will include the other \textsc{Fastsum} ensembles, Generation 2~\cite{Aarts:2014nba,Aarts:2020vyb,aarts_2023_8403827} and 2L~\cite{Aarts:2020vyb,Aarts:2022krz,aarts_2024_10636046}, to give an estimate of systematic effects associated with the pion mass and the anisotropy. It is interesting that for our ensembles the effective restoration temperature for $U(1)_A$ is close to the transition temperature found in \Refl{Mickley:2024vkm} via centre vortices, indicating the possible existence of a third high-temperature phase of QCD matter~\cite{Alexandru:2019gdm,Mickley:2024vkm,Kotov:2025ilm}. We also look forward to applying these methods to the puzzle of chiral spin symmetry~\cite{Rohrhofer:2019qwq,Chiu:2024bqx,Glozman:2025twe,Philipsen:2025lda}.
%\vspace{-0.015\textheight}
\section*{Software and Data}
%\vspace{-0.015\textheight}
The Generation 3 ensembles were generated using \textsc{OpenQCD-Fastsum}~\cite{glesaaen_2018_2216356}, a derivative of \textsc{OpenQCD-1.6}. They will be made available at some point in the future in accordance with \textsc{Fastsum}'s sharing policy. The correlators and analysis workflow will be made available along with a future publication. This analysis makes extensive use of the \textsc{python} packages \textsc{gvar}~\cite{peter_lepage_2025_14783421}, \textsc{matplotlib} and \textsc{NumPy}. Error analysis is performed through a combination of \textsc{gvar} and a jackknife analysis implemented in \textsc{Fortran} using the \textsc{Fortran-Package-Manager}~\cite{DBLP:journals/corr/abs-2109-07382,779aad0a0cba4c0297f31b532bd4aca7} with \textsc{python} bindings.~\cite{fortran_meson}.
%\vspace{-0.015\textheight}
\section*{Acknowledgements}
%\vspace{-0.015\textheight}
This work is supported by STFC grant ST/X000648/1. GA is supported by a Royal Society Leverhulme Trust Senior Research Fellowship. RB acknowledges support from a Science Foundation Ireland Frontiers for the Future Project award with grant number SFI-21/FFP-P/10186. SK is supported by the National Research Foundation of Korea through the grant, NRF-2008-000458. We acknowledge the EuroHPC Joint Undertaking for awarding the projects EHPC-EXT-2023E01-010 and EXT-2025E01-079 access to LUMI-C and LUMI-G, Finland. This work used the DiRAC Data Intensive service (DIaL2 \& DIaL) at the University of Leicester, managed by the University of Leicester Research Computing Service on behalf of the STFC DiRAC HPC Facility (www.dirac.ac.uk). The DiRAC service at Leicester was funded by BEIS, UKRI and STFC capital funding and STFC operations grants. This work used the DiRAC Extreme Scaling service (Tesseract) at the University of Edinburgh, managed by the Edinburgh Parallel Computing Centre on behalf of the STFC DiRAC HPC Facility (www.dirac.ac.uk). The DiRAC service at Edinburgh was funded by BEIS, UKRI and STFC capital funding and STFC operations grants. This work was performed using PRACE resources at Joliot-Curie (Irene) hosted in Rome and Hawk hosted by HLRS Stuttgart. We acknowledge the support of the Supercomputing Wales project, which is part-funded by the European Regional Development Fund (ERDF) via Welsh Government. 
%\vspace{-0.015\textheight}
\bibliographystyle{JHEP}
\bibliography{skeleton}% Produces the bibliography via BibTeX.

\end{document}